\documentclass[twocolumn,secnumarabic,amssymb, showpacs,amsmath, nofootinbib,tightenlines,
  nobibnotes, prl]{revtex4}

\usepackage{graphicx}
\usepackage{dcolumn}
\usepackage{bm}
\usepackage{multirow}

\begin{document}

\preprint{APS/123-QED}

\title{Attosecond Magnetic Field Pulse Generation by Intense Few Cycle Circularly Polarized UV Pulses  }

\author{Kai-Jun Yuan}
\author{Andr\'{e} D. Bandrauk}%
 \email{andre.bandrauk@usherbrooke.ca}
\affiliation{%
D\'{e}partement de Chimie, Facult\'{e} des
Sciences, Universit\'{e} de Sherbrooke, Sherbrooke, Qu\'{e}bec,
Canada, J1K 2R1
}%

\date{\today}

\begin{abstract}
Intense attosecond magnetic field pulses are predicted to be produced by intense few cycle attosecond circularly polarized UV pulses. Numerical solutions of the time dependent Schr\"{o}dinger equation for H$_2^+$ are used to study the electronic dynamical process. Spinning attosecond circular electron wave packets are created on sub-nanometer molecular dimensions, thus generating attosecond magnetic fields of several tens of Teslas ($10^5$ Gauss). Simulations show that the induced magnetic field is critically dependent on the pulse wavelength $\lambda$ and pulse duration $n\tau$ ($n$ number of cycle) as predicted by a classical model. For ultrashort few cycle circularly polarized attosecond pulses, molecular orientation influences the generation of the induced magnetic fields as a result of preferential ionization perpendicular to the molecular axis. The nonspherical asymmetry of molecules allows for efficient attosecond magnetic field pulse generation.

\end{abstract}

\maketitle
\section{introduction}
The rapid developments in synthesizing ultrashort intense pulses offer the possibility to investigate electron dynamics on its natural time scale, the attosecond (1 as=$10^{-18}$ s) in the nonlinear nonperturbative regime of laser molecule interaction \cite{NP:,RMP:81:163,changcorkum}. To date the shortest linearly polarized single pulse with duration of 67 as has been produced from high-order harmonic generation (HHG) with a few cycle intense infrared laser field in atoms \cite{ol:chang}. With a mid-infrared femtosecond laser HHG spectra of very high order $\sim 5000$ (1.6 keV)
can also be generated thus allowing for the generation of pulses
as short as few attoseconds \cite{sci:jila}. With such attosecond pulses electrons in matter can then be visualized and controlled on the attosecond time scale and
sub-nanometer dimension, e.g., \cite{JPB:39:S409,PRL:94:083003,PRL:105:263201,marc}.  An application of such pulses is the monitoring of coherent superposition of two or more bound electronic states in atomic or molecular systems
which lead to electron motion on attosecond time scale with momentum distributions \cite{PRL:} or HHG spectra \cite{STA} measurable as a
function of the time delay between pump-probe pulses. However the development of attosecond pulse technology has been limited to linear polarization. Circularly polarized attosecond pulses have been now proposed as future tools for studying further attosecond electron dynamics \cite{Eberly,Uzer,PRL:99:047601,14}. We have proposed previously methods of generating such {circularly} polarized attosecond pulses from circularly polarized molecular HHG due to the nonsymmetry of molecular Coulomb potentials \cite{yuan:prl}. Atomic Coulomb potentials can also control recollision of electrons in double ionization with circular polarization for specific atomic parameters \cite{Uzer}. 

\begin{figure}[!t]\centering
\includegraphics[scale=0.95,angle=0]{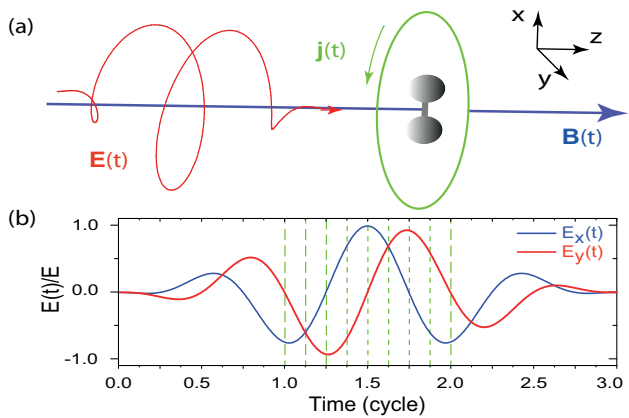}
\caption{(a) Illustration of the attosecond magnetic fields $\textbf{B}(\textbf{r},t)$ (blue line, along the $z$ axis) for H$^+_2$ by a few cycle circularly polarized attosecond UV pulse (red line). The green line represents the corresponding current $\textbf{j}(\textbf{r},t)$ in the molecular ($x,y$) plane. The magnetic field $\textbf{B}(\textbf{r},t)$ is perpendicular to the current $\textbf{j}(\textbf{r},t)$. (b) Three cycle circularly polarized attosecond UV pulse $\textbf{E}(t)$ at $\lambda=50$ nm. Dashed lines correspond to the times in Figs. \ref{Fig1} and \ref{Fig2}, and Table \ref{tab1}.  }
\label{Fig0}
\end{figure}

Optically induced ultrafast magnetization reversal has now been reported by circularly polarized pulses in material science \cite{PRL:99:047601}, thus emphasizing new and future application of such pulses. Optically induced magnetic fields have been previously related to the inverse Faraday effect \cite{IFE}, for which a perturbative weak magnetic field treatment has been presented in terms of optical polarizabilities \cite{IFE2}. Recently, the generation of electronic ring currents in aromatic molecules obtained from quantum-chemical numerical simulations can produce static magnetic fields by means of linear \cite{PRL:lin} and circularly polarized $\pi$ UV pulses resonant with degenerate $\pi$ orbitals \cite{manz:JACS,21}. The generated magnetic fields can be much larger than those obtained by traditional static fields \cite{19}. The driving pulses can be optimized as well by optimal control theory \cite{PRL:gross}. In these previous studies, coherent rotational electronic states are prepared resonantly, thus leading to static magnetic fields. Moreover these induced effects strongly rely on the coherence of the excited states. With intense circularly polarized attosecond pulses, we have proposed to create ``spinning" continuum electrons which can be produced and localized on the sub-nanometer molecular dimensional scale \cite{yuan:jpb}. As a result, circular electron wave packets and currents are created in the continuum and thus large internal molecular magnetic fields are generated. In this paper we show that intense attosecond magnetic field pulses can be obtained from intense few cycle circularly polarized attosecond UV pulses, as illustrated in Fig. \ref{Fig0}. The developments of circularly polarized attosecond pulses \cite{yuan:prl} make it thus possible to create spinning circular electron wave packets, leading to large time-dependent internal magnetic fields in matter on attosecond time scale.

   The paper is arranged as follows: In Sec. 2, we briefly
describe the computational methods for time-dependent quantum
electron wave packets for calculations from the
 corresponding TDSEs. The numerical results of time-dependent electronic currents and magnetic fields
 by intense ultrashort few cycle circularly polarized UV pulses for H$_2^+$ are
presented in Sec.
3. Effects of the pulse wavelength and molecular orientation are also presented. Finally, we summarize our findings in Sec.
4. Throughout this paper, atomic units (a.u.) $e=\hbar=m_e=1$
are used unless otherwise stated.

\section{numerical methods} \label{methods}

Simulations are performed on an oriented fixed nuclei molecular ion H$_2^+$ from numerical solutions of the corresponding time dependent Schr\"{o}dinger equation (TDSE) within a static (Born-Oppenheimer approximation, BOA) nuclear frame,  \begin{equation} i\frac{\partial}{\partial t}\psi(\textbf{r},t)=[\hat H_0+V_L({\textbf{r}})]\psi(\textbf{r},t),\end{equation}  where $H_0$ is the field free molecular Hamiltonian and $V_L({\textbf{r}})$ the interaction term. Such a fixed nuclei approach by ignoring the vibrational and rotational degrees of freedom is appropriate due to the longer femtosecond/picosecond time scale of nuclear (vibrational/rotational)
motions. The corresponding three dimensional (3D) TDSE is solved using cylindrical
coordinates $\textbf{r}=(\rho,\theta,z$) with the molecular plane $x=\rho\cos\theta$ and $y=\rho\sin\theta$, Fig. 1(a). The molecular Hamiltonian is expressed as
\begin{eqnarray}H_0(\rho,\theta,z)&=&-\frac{1}{2\rho}\frac{\partial}{\partial \rho}\left(\rho\frac{\partial}{\partial \rho}\right)-\frac{1}{2\rho^2}\frac{\partial^2}{\partial \theta^2} \nonumber \\ && -\frac{1}{2}\frac{\partial^2}{\partial z^2}+V(\rho,\theta,z), \end{eqnarray} where $V(\rho,\theta,z)$ is the two center electron-nuclear potential. The molecular ion is pre-oriented before ionization and this can be readily achieved with orientational laser technology \cite{3d:alignment}. The radiative interaction between the laser field and the electron is described by \begin{equation}V_L(\textbf{r})=\textbf{r}\cdot \textbf{E}(t)=\hat e_x \rho\cos\theta E_x(t)+\hat e_y\rho\sin\theta E_y(t)\end{equation}
in the length gauge for circularly polarized pulses,
$\textbf{E}(t)=E f(t)[\hat e_x\cos(\omega t)+ \hat e_y
\sin(\omega t)],$ propagating in the $z$ direction and
 $\hat e_{x/y}$ is the polarization direction. A smooth $\sin^2(\pi t /n\tau)$ pulse envelope $f(t)$
for maximum
amplitude $E $ and intensity $I =I_x+I_y={ c\varepsilon_0
E ^2}$ is adopted, where one optical cycle $\tau=2 \pi/\omega$. This pulse satisfies the total zero area $\int E(t) dt=0$ in order to exclude static field effects [26-27].

The 3D TDSE in Eq. (1) is solved numerically by a
second order split operator method \cite{JCP:99:1185} in the time step $\delta t$
combined with a fifth order finite difference method and Fourier
transform technique in the spatial steps $\delta \rho$, $\delta z$, and $\delta
\theta$ \cite{Lubook}. The time
step is taken to be $\delta t=0.01$ a.u.=0.24 as. The spatial
discretization is $\delta \rho=\delta z=0.25$ a.u. for a radial grid range
$0 \leq \rho \leq 128$ a.u. (6.77 nm) and $|z|\leq$ 32 a.u. (1.69 nm), and the angle grid size $\delta
\theta=0.025$ radian. To prevent unphysical effects due to the
reflection of the wave packet from the boundary, we multiply
$\psi(\rho,\theta,z,t)$ by a ``mask function" or absorber in the radial coordinates $\rho$
with the form $\cos^{1/8}[\pi(\rho-\rho_\textrm{a})/2\rho_{\textrm{abs}}]$. For all results reported here we set the absorber domain
$\rho_{\textrm{a}}=\rho_{\textrm{max}}-\rho_{\textrm{abs}}$=104 a.u. with $\rho_{\textrm{abs}}=24$
a.u., exceeding well the field induced electron oscillation
$\alpha_d={E }/{\omega^2}$ of the electron.

The time dependent electronic current density is defined by the quantum expression,
\begin{equation} \label{current}
 \textbf{j}(\textbf{r},t)=\frac{i}{2}[\psi(\textbf{r},t) \nabla_{\textbf{r}} \psi^*(\textbf{r},t) -\psi^*(\textbf{r},t)\nabla_{\textbf{r}}\psi(\textbf{r},t)],
\end{equation} $\psi(\textbf{r},t)$ is the exact BOA electron wave function obtained from the TDSE and $\nabla_{\textbf{r}}=\textbf{e}_\rho\nabla_\rho+\textbf{e}_\theta\frac{1}{\rho}\nabla_\theta+\textbf{e}_z\nabla_z$ in cylindrical coordinates. Then the corresponding time dependent magnetic field is calculated using the following classical Jefimenko's equation \cite{Jefbook}
\begin{equation} \label{magnetic}
 \textbf{B}(\textbf{r},t)=\frac{\mu_0}{4\pi}\int [\frac{\textbf{j}(\textbf{r}',t_r)}{|\textbf{r}-\textbf{r}'|^3}+\frac{1}{|\textbf{r}-\textbf{r}'|^2c}\frac{\partial \textbf{j}(\textbf{r}',t_r)}{\partial t}]\times (\textbf{r}-\textbf{r}')d^3\textbf{r}',
\end{equation}
where $t_r=t-r/c$ is the retarded time and $\mu_0=4\pi \times 10^{-7}$ NA$^{-2}$ (6.692$\times10^{-4}$ a.u.). Units of $B(\textbf{r},t)$ are Teslas (1T=$10^{4}$ Gauss). For the static time-independent conditions occurring after the pulse, then Eq. (\ref{magnetic}) reduces to the classical Biot-Savart law \cite{Jefbook}.

\begin{figure}[!t]\centering
\includegraphics[scale=.9,angle=0]{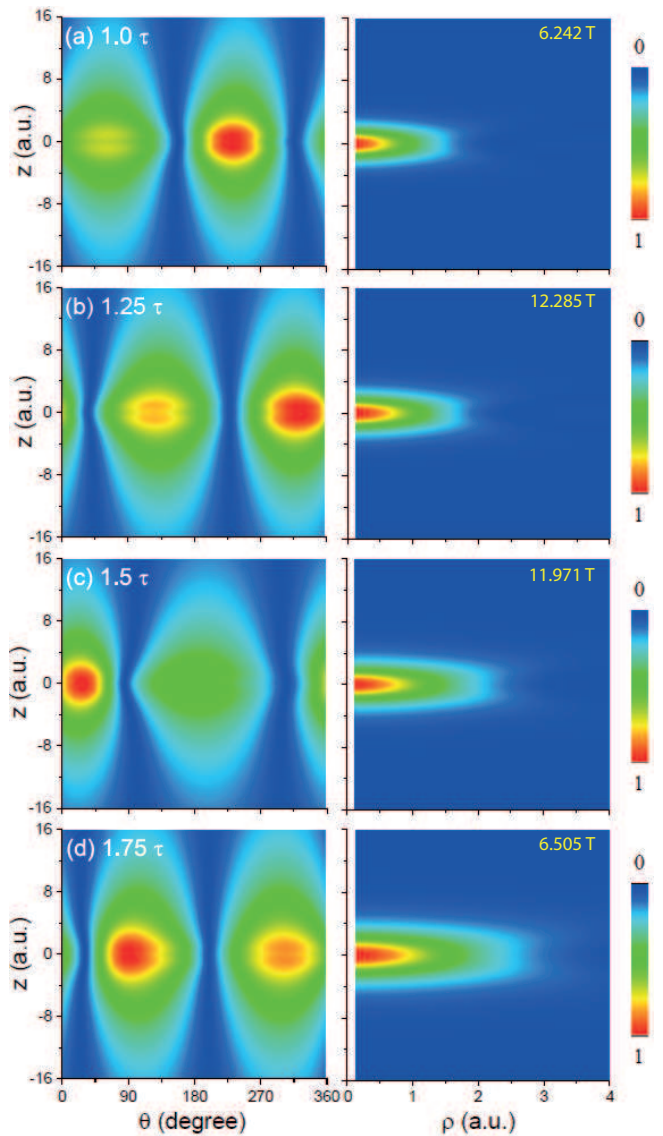}
\caption{ Attosecond magnetic fields $|\textbf{B}(z,\theta,t)|$ ($\theta=\tan^{-1}(y/x)$) (left column) and $|\textbf{B}(z,\rho,t)|$ (right column) at different moments generated by circularly polarized attosecond UV pulses, illustrated in Fig. \ref{Fig0} with intensity $I=2\times10^{16}$ W/cm$^2$ ($E=0.5338$ a.u.), wavelength $\lambda=50$ nm ($\omega=0.911$ a.u.), and duration $3\tau=500$ as. The corresponding values of the magnetic field strengths (T) are listed in Table \ref{tab1}. The corresponding maximum local magnetic fields $B_{max}(\textbf{r},t)$ are respectively (a) 6.242 T at $t=1.0\tau$, (b) $12.285$ T at $t=1.25\tau$, (c) $11.971$ T at $t=1.5\tau$, and (d) $6.505$ T at $t=1.75\tau$.}
\label{Fig1}
\end{figure}

\begin{figure}[!t]\centering
\includegraphics[scale=0.9,angle=0]{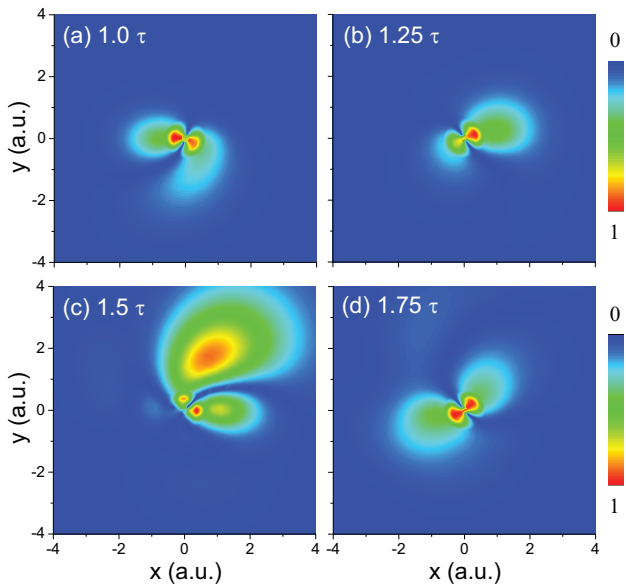}
\caption{ Evolution of current densities $|\textbf{j}(x,y,t)|$ (fs$^{-1}\cdot a_0^{-1}$) with time $t$ generated by circularly polarized attosecond UV pulses with intensity $I=2\times10^{16}$ W/cm$^2$, wavelength $\lambda=50$ nm, and duration $3\tau=500$ as, corresponding to Fig. \ref{Fig1}. The corresponding values of the currents are listed in Table \ref{tab1}.}
\label{Fig2}
\end{figure}

\section{results and discussions}

In the present simulations, the molecule is initially in the ground $1s\sigma_g$ state of H$_2^+$ at equilibrium $R_e=2$ a.u. with the ionization potential $I_p=1.1$ a.u. (29.72 eV). The initial corresponding wavefunction $\psi(\textbf{r},t=0)$ is obtained by propagating an
initial appropriate wavefunctions in imaginary time
using the zero-field TDSE [26,27]. We adopt a circularly polarized attosecond UV pulse with intensity $I=2\times10^{16}$ W/cm$^2$ ($E=0.5338$ a.u.), corresponding to $I_x=I_y=1\times10^{16}$ W/cm$^2$, wavelength $\lambda=50$ nm ($\omega=0.911$ a.u.), and duration $3\tau=500$ as. With such pulses of broad spectral width $\sim0.44$ a.u.=11.97 eV, a single photon absorption is possible releasing the electron with near zero initial velocity $v(t_0)\approx0$. The Keldysh parameter $\gamma=\sqrt{I_p/2U_p}=2.5$, where $U_p=E^2/4\omega^2$ is the ponderomotive energy, indicates a region of multi-photon ionization processes [30,31].

For the $x$ oriented H$_2^+$ ionized by circularly polarized attosecond UV pulses propagating along the $z$ axis (Fig. \ref{Fig0}), the current densities $|\textbf{j}(\textbf{r},t)|$ (fs$^{-1}$) are mainly localized in the molecular plane ($x,y$) and the corresponding magnetic fields $\textbf{B}(\textbf{r},t)$ (T) are along the $z$ axis, perpendicular to the molecule, as predicted in Eqs. (\ref{current}) and (\ref{magnetic}) and illustrated in Fig. \ref{Fig0}. We illustrate in Fig. \ref{Fig1} snapshots of the induced attosecond magnetic fields $\textbf{B}(z,\theta,t)$ averaged over $\rho$ (left column) as a function of angle $\theta$ in the $(x,y)$ molecular plane and $\textbf{B}(z,\rho,t)$ averaged over $\theta$ (right column) as a function of radial coordinates at different moments around the peak time $t=1\sim2\tau$ of the pulse. The magnetic fields evolve around the $z$ axis in the ($x,y$) plane due to the circularly polarized pulses. There are two field components which are symmetric with respect to the molecular ($x,y$) plane and centered at $z\approx\pm1$ a.u. (0.0529 nm) above and below the molecular plane. The magnetic fields rotate from angles $\theta=-90^\circ$ (270$^\circ$) to $90^\circ$ with time, and at $1.25\tau \leq t\leq1.5\tau$, the field is maximum at $\theta=360^\circ$ (0$^\circ$) parallel to the molecular $x$ or $R$ axis. The corresponding current densities $|\textbf{j}(x,y,t)|$ obtained by integrating over $z$ perpendicular to the molecular plane are also displayed in Fig. \ref{Fig2}. From Figs. \ref{Fig1} and \ref{Fig2} one sees that both $|\textbf{j}(x,y,t)|$ and $\textbf{B}(z,\theta,t)$ spirally rotate with an anti-clock wise direction, following the left-handed polarization.

For a circularly polarized pulse with wavelength $\lambda=50$ nm below the ionization threshold and duration $3\tau=500$ as, the corresponding spectral width at half maximum is about $\Delta\omega=0.44$ a.u.. After absorption of one photon of frequency $\omega=0.911$ a.u., electron wave packets are created in the continuum with relatively small kinetic energy with dominant distributions near zero initial velocity. The electron wave packets then move under the influence of the intense pulse. As a result, the current appears in the molecular ($x,y$) plane perpendicular to the propagation $z$ direction. This time dependent current produces a dynamical attosecond magnetic field which is a function of the pulse phase $\omega t$.

\begin{table*}[!btp] \centering
\caption{\label{tab1} Maximum local $B_{max}(\textbf{r},t)$ and total volume magnetic field ${B}(t)$ and currents $j(t)$ by circularly polarized attosecond UV pulses with intensity $I=2\times10^{16}$ W/cm$^2$, wavelength $\lambda=50$ nm, and duration $3\tau=500$ as at different times, c.f. Figs. \ref{Fig1} and \ref{Fig2}.}
\begin{tabular} {cccccccc}
\hline \hline
& units &1.0$\tau$ & 1.25$\tau$ &1.5$\tau$ &1.625$\tau$& 1.75$\tau$& 2.0$\tau$\\ \hline

\multirow{2}{*}{$B_{max}(\textbf{r},t)$} & a.u.($\times 10^{-4}$) &0.250&0.491 & 0.479   & 0.365  &0.260  &0.167  \\
 & T & 6.242 &  12.285 & 11.971 & 9.121 & 6.505& 4.167  \\

 \multirow{2}{*}{$B(t)$} & a.u.($\times 10^{-4}$) &0.267&0.705 &  1.112 & 1.172 &1.101 & 0.693\\
 & T$\cdot a_0^3$ & 6.670 &  17.612 & 27.797&29.300 & 27.517& 17.320 \\

\multirow{3}{*}{$j(t)$} & a.u. &0.126& 0.304 &  0.257 & 0.247&0.249 &0.202\\
  & fs$^{-1}$$\cdot a_0$ & 5.209 & 12.568 & 10.625 & 10.211 &10.294 &8.351\\
  & mA$\cdot a_0$ & 0.834 &  2.011 &  1.700 & 1.634 &1.674 & 1.336 \\
 \hline\hline
\end{tabular}
\end{table*}

We calculate the maximum local $B_{max}(\textbf{r},t)$ and total volume average attosecond magnetic fields and the corresponding currents by integrating $B(t)=|\int\textbf{B}(\textbf{r},t)d\textbf{r}^3|$ and $j(t)=|\int\textbf{j}(\textbf{r},t)d\textbf{r}^3|$ ($d\textbf{r}^3=\rho d\rho d\theta dz$) over the electron $\textbf{r}$ space. Table \ref{tab1} lists values of $B_{max}(\textbf{r},t)$, $B(t)$, and  $j(t)$ at different moments, illustrated in Figs. \ref{Fig1} and \ref{Fig2}. Both $B(t)$ (in units of T$\cdot a_0^3$, where $a_0$ is Bohr radius) and $j(t)$ (fs$^{-1}\cdot a_0$) vary with time, increasing first and then decreasing in phase with the pulse. In Figs. 2 and 3 we see that the induced volume electronic currents and the attosecond volume magnetic field pulses are localized and concentrated in space. From Tab. \ref{tab1} one obtains that the maximum total volume magnetic field is induced at $t=1.625\tau$=270 as with strength $B=1.172\times10^{-4}$ a.u. =29.3 T$\cdot a_0^3$ (2.93$\times10^5$ Gauss$\cdot a_0^3$). The maximum local magnetic field $B_{max}(\textbf{r},t)=12.285$ T and the maximum electronic current $j=0.304$ a.u.=12.568 fs$^{-1}$$\cdot a_0$ (2.011 mA$\cdot a_0$) is produced at time $t=1.25\tau=207$ as, where $E_x=0$ and $E_y=-E$ (Fig. 1).
A time delay $\Delta t=0.375\tau=63$ as occurs between maximum current $j(t=1.25\tau)=12.568$ fs$^{-1}$$\cdot a_0$ and $B(t=1.625\tau)=29.3$ T$\cdot a_0^3$ (Table \ref{tab1}). As shown in Fig. \ref{Fig0}, the ionization mainly occurs at $1.125\tau \leq t\leq1.25\tau$ where $E_y$ is maximum due to the nonspherical molecular Coulomb potential \cite{yuan:jpb2}. Thus the maximum current is obtained as listed in Table \ref{tab1}. After ionization, the electron moves following the electric fields, $E_x>0$ and $E_y<0$, leading the rotation trajectory with anti-clock wise direction, as illustrated in Fig. \ref{Fig2}(b,c). At time $t=1.625\tau$, the maximum total magnetic field is produced. From Fig. \ref{Fig2} one then observes the magnetic field angle is near $90^\circ$, corresponding to the $y$ axis, essentially perpendicular to the molecular $R$ axis.

\begin{figure}[!t]\centering
\includegraphics[scale=0.6,angle=0]{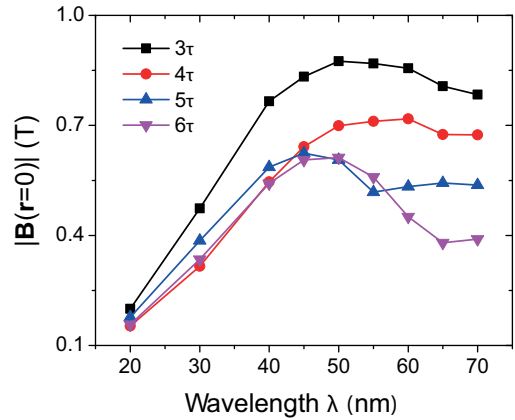}
\caption{ Maximum values of the induced local magnetic fields $|\textbf{B}(\textbf{r}=0)|$ (T) at the central point $\textbf{r}=0$ by $I =2\times10^{16}$ W/cm$^2$ circularly polarized UV laser pulses with different wavelengths $\lambda$ and durations $n\tau$ (cycles).}
\label{Fig3}
\end{figure}

We next show effects of the pulse wavelength $\lambda$ and pulse duration $n\tau$ on the magnetic field. The maximum values of the local magnetic field strength $\textbf{B}(\textbf{r}=0)$ at $\textbf{r}=0$, the center of the molecule are displayed in Fig. \ref{Fig3}. We fix the pulse intensity at $I=2\times10^{16}$ W/cm$^2$. The pulse wavelength is varied from $\lambda=20$ nm to 70 nm for different pulse durations $n\tau$, where $n=3,\cdot\cdot\cdot,6$ is the number of cycles. $\textbf{B}(\textbf{r}=0)$ strongly depends on these pulse parameters. At short wavelength $\lambda=20$ nm weak $|\textbf{B}(\textbf{r}=0)|\approx0.17$ T are induced, which however are insensitive to the pulse duration $n\tau$. Increasing $\lambda$ the corresponding magnetic field increases gradually. One sees that the results for the $3\tau$ pulses vary quickly. At $\lambda=50$ nm the magnetic field reaches a maximum strength $|\textbf{B}(\textbf{r}=0)|=0.88$ T, five times stronger than that at $\lambda=20$ nm. For the case of $n\geq4$ cycles, the magnetic field is nearly insensitive to the pulses duration. Increasing the pulse wavelength $\lambda$ or decreasing the frequency $\omega$ further, the magnetic field decreases since the induced electron radius is inversely proportional to the frequency $\omega$ as described next. Moreover, as the pulse duration $n\tau$ increases $\textbf{B}(\textbf{r}=0)$ decreases dramatically. At $\lambda=70$ nm, the magnetic field strength of $3\tau$ is twice that of $6\tau$.

At shorter pulse wavelengths $\omega>I_p$ or higher frequencies, the molecule is ionized after absorption of one photon and the corresponding continuum electron with initial velocity $\sqrt{2(\omega-I_p)}$ moves away from the molecular center directly but mainly perpendicular to the molecular $R$ axis, i.e., the $y$ direction \cite{yuan:jpb2} due to a nonzero drift velocity in the $y$ direction \cite{29}. Furthermore ionization rates are also weak. Weaker magnetic fields are obtained which are nearly insensitive to the pulse duration $\tau$. For the lower pulse frequency $\omega \lesssim I_p$, the electron is released from the molecules with low initial velocities by absorbing one photon due to the broad spectral width of pulses. The electron trajectories are functions of the pulse wavelength $\lambda$ (frequency $\omega$) and duration $n\tau$ \cite{yuan:jpb}. We explain this from the classical model \cite{yuan:jpb}, a generalization of the linear polarization model \cite{3-step}. Assuming the zero initial electron velocities $\dot x(t_0)=\dot y(t_0)=0$, where $t_0$ is the ionization time, the induced time dependent velocities are \begin{eqnarray} \label{VC} \displaystyle { \begin{array}{l}
\dot x(t)=\displaystyle -{E}/{\omega} \left( \sin\omega
t-\sin\omega t_0 \right ),  \\ \dot y(t)=\displaystyle
-{E}/{\omega} \left( \cos\omega t_0 -\cos \omega t
 \right ),\end{array} }
\end{eqnarray}
The corresponding displacements
are
\begin{eqnarray} \label{DC} \displaystyle { \begin{array}{l}
 x(t)=\displaystyle -{E}/{\omega^2} \left[\cos\omega t_0-\cos\omega t-(\omega t-\omega t_0)\sin\omega t_0 \right ],
\\y(t)=\displaystyle -{E}/{\omega^2} \left[\sin\omega
t_0-\sin\omega t+ (\omega t-\omega t_0)\cos\omega t_0
 \right ],\end{array} }
\end{eqnarray}
with $x(t_0)=y(t_0)=0$ corresponding to recollision at the center of the molecule ($r=0$). From Eqs. (\ref{VC}) and (\ref{DC}) it is found that increasing the pulse wavelength $\lambda$ leads to increase of the maximum induced electron velocity $v=2E/\omega$ at $\omega
t-\omega t_0=(2n'+1)\pi$, and the corresponding radii \begin{equation} r_{n'}={2E}/{\omega^2} \left [
1+(n'+{1}/{2})^2\pi^2 \right ]^{1/2},\end{equation} $n'=0,1,2,\cdots$. For a moving point charge the corresponding classical magnetic field can be expressed as \begin{equation} \label{BC}
\textbf{B}=\frac{\mu_0}{4\pi}\frac{\textbf{v}\times\textbf{r}}{\textbf{r}^3}.\end{equation} From Eqs. (\ref{VC}), (\ref{DC}) and (\ref{BC}) one then gets the maximum field $B\sim1/r_{n'}^2$ at times $t_0+(2n'+1)\pi/\omega$.
Therefore, an increase of $\lambda$ or lower $\omega$, results in a decrease of the magnetic field due to large radii $r_{n'}$ of the electron, thus reducing the efficiency of the attosecond magnetic field generation. Longer pulse duration have a similar effects. Of note is that we only present the results for the single photon ionization processes. Increasing the pulse wavelength further, more photons are required to ionize molecules, resulting in decrease of the ionization rate and increase of the radii of the free photoelectron. Consequently, weaker magnetic field pulses are produced, see Eq. (9).

\begin{figure}[!t]\centering
\includegraphics[scale=0.6,angle=0]{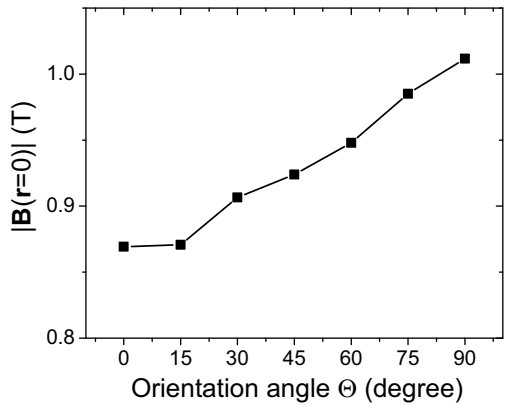}
\caption{Maximum values of the laser induced local magnetic fields $|\textbf{B}(\textbf{r}=0)|$ (T) at $\textbf{r}=0$ by intense circularly polarized UV laser pulses with intensity $I =2\times10^{16}$, wavelengths $\lambda=50$ nm and durations $3\tau=500$ as for different orientation angles $\Theta$ between the molecular $R$ axis and the $E_x$ field component.}
\label{Fig4}
\end{figure}

For few cycle circularly polarized attosecond pulses, the molecular orientation can also influence the magnetic field. In Fig. \ref{Fig4} we display the maximum values of the local $|\textbf{B}(\textbf{r}=0)|$ for different angles $\Theta$, where $\Theta$ is the molecular orientation angle between the molecular $R$ axis and the $x$ axis, the direction of the $E_x$ component. $|\textbf{B}(\textbf{r}=0)|$ increase as $\Theta$ increases. The same pulse is used as in Fig. \ref{Fig1}. For the three cycle circularly polarized attosecond pulse, maxima of the $E_x(t)$ and $E_y(t)$ field components are not equal. Moreover, due to the nonspherical molecular Coulomb potential the ionization ratio between the parallel $x$ and the perpendicular $y$ polarizations is unequal and the ionization probability is dominant perpendicular to the molecular $R$ axis due to two center interference \cite{yuan:jpb2} and perpendicular drift \cite{29}. Since the ionization probability is sensitive to $\Theta$, a change of the molecular orientation influences the current as well as the magnetic field. As illustrated in Fig. \ref{Fig0}(b) the maximum of the field component $E_x(t=1.5\tau)=E$ is slightly stronger than that of $E_y(t=1.25\tau)=0.93E$ for the molecular ionization. The largest magnetic field is then obtained at $\Theta=90^\circ$ with $|\textbf{B}(\textbf{r}=0)|$=1.01 T.


\section{Conclusions}

In summary, we have theoretically presented generation of intense dynamical attosecond magnetic fields in the molecular ion H$_2^+$ using few cycle circularly polarized attosecond UV pulses by creating time-dependent electron ring currents in the continuum. Comparing with those previous processes in aromatic molecules where the electric currents are created by stationary state electron currents of $\pi$ orbitals [18-20], one sees that generation of the attosecond magnetic field pulses from free electron currents can be readily controlled to yield magnetic fields $\textbf{B}(t)$ which are a function of velocities ${\bf v}$ and displacements $\textbf{r}$ of ionized electrons, in Eq. (9). These are determined by frequency and duration of circularly polarized attosecond pulses [23]. Using a circularly polarized attosecond pulse with proper conditions, strong magnetic field pulses with several tens of Teslas can be produced. The radii of the free electron is critically sensitive to the pulse wavelength as defined by the classical model, Eq. (7). As a result altering the pulse wavelength leads to increase or decrease of the efficiency of the attosecond magnetic field pulse generation. Such scheme to generate attosecond magnetic field pulses can also be extend to complex molecular systems with many nuclear centers and multiple electrons. In the latter case, inner shell ionization must be considered \cite{xx}. Of note is that interference effects of multiple electronic currents will influence the attosecond magnetic field pulse generation. These magnetic field pulses should be useful for studying molecular paramagnetic bonding \cite{PRX}, nonequilibrium electronic processes \cite{y}, demagnetization processes \cite{z}, and optical magnetic recording \cite{PRL:99:047601}, thus offering experimentalists new tools for controlling electron dynamics on attosecond time scale.
\section*{ACKNOWLEDGEMENTS}
The authors thank RQCHP and Compute Canada for access to massively parallel
computer clusters and CIPI (Canadian Institute for Photonic
Innovations) for financial support of this research in its
ultrafast science program.
\providecommand{\newblock}{}

\end{document}